\documentclass[prd,aps,twocolumn,showpacs, nofootinbib,superscriptaddress,notitlepage]{revtex4-1}
\usepackage{amsmath,graphicx,epsfig,amssymb}
\usepackage{inputenc}
\usepackage[usenames]{color}
\usepackage{slashed}
\usepackage{multirow}
\usepackage{subfigure}
\usepackage{dsfont}
\usepackage{comment}
\usepackage{steinmetz}
\usepackage{setspace}

\allowdisplaybreaks

\usepackage{floatrow}
\newfloatcommand{capbtabbox}{table}[][\FBwidth]

\begin{document}
\title{QED contributions to  the $\Xi_c-\Xi_c'$ mixing}

\author{Zhi-Fu Deng}
\affiliation{INPAC, Key Laboratory for Particle Astrophysics and Cosmology (MOE),  Shanghai Key Laboratory for Particle Physics and Cosmology, School of Physics and Astronomy, Shanghai Jiao Tong University, Shanghai 200240, China}
\author{Yu-Ji Shi}\email{Corresponding author: shiyuji@ecust.edu.cn}
\affiliation{School of Physics, East China University of Science and Technology, Shanghai 200237, China}
\affiliation{INPAC, Key Laboratory for Particle Astrophysics and Cosmology (MOE),  Shanghai Key Laboratory for Particle Physics and Cosmology, School of Physics and Astronomy, Shanghai Jiao Tong University, Shanghai 200240, China}
\author{Wei Wang}  
\affiliation{INPAC, Key Laboratory for Particle Astrophysics and Cosmology (MOE),  Shanghai Key Laboratory for Particle Physics and Cosmology, School of Physics and Astronomy, Shanghai Jiao Tong University, Shanghai 200240, China}
\affiliation{Southern Center for Nuclear-Science Theory (SCNT), Institute of Modern Physics, Chinese Academy of Sciences, Huizhou 516000, Guangdong Province, China}
\author{Jun Zeng}\email{Corresponding author: zengj@sjtu.edu.cn }
\affiliation{INPAC, Key Laboratory for Particle Astrophysics and Cosmology (MOE),  Shanghai Key Laboratory for Particle Physics and Cosmology, School of Physics and Astronomy, Shanghai Jiao Tong University, Shanghai 200240, China}

\date{\today}

\begin{abstract}
We explore the QED corrections to the $\Xi_c-\Xi_c^{\prime}$ mixing  within the framework of light-front quark model (LFQM) in the three-quark picture. After explicitly investigating the relation between the $\Xi_c-\Xi_c^{\prime}$ mixing and the flavor $\rm {SU(3)}$ and heavy quark symmetry breaking, we derive the QED contributions to the mixing angle.  Numerical results indicate the QED contribution is smaller than the one from the mass difference between the strange and up/down quark provided by a recent Lattice QCD analysis. Adding these contributions together we find that at this stage  the $\Xi_c-\Xi_c^{\prime}$ mixing is small and still incapable to account for the large $\rm {SU(3)}$ symmetry breaking in the semi-leptonic $\Xi_c$ decays.
\end{abstract}

\maketitle

\section{Introduction}
\label{sec:introduction}
Weak decays of heavy baryons provide an ideal platform for understanding the strong interaction and searching for new physics beyond the standard model. One of the widely used theoretical methods in this area is the flavor $SU(3)$ symmetry, where the $u, d, s$ quark are treated as identical \cite{Zeppenfeld:1980ex,Savage:1989ub,Deshpande:1994ii,He:1998rq,Chau:1987tk,Gronau:1994rj,Gronau:1995hm,Cheng:2014rfa,Muller:2015lua,Shi:2017dto}. Using the flavor $SU(3)$ symmetry one can classify the singly heavy baryons into anti-triplet and sextet, with the two light quarks inside the baryon forming a scalar and an axial-vector, respectively. However, since the $u, d, s$ quarks have different masses and electric charges, the flavor $SU(3)$ symmetry is broken. Therefore, the anti-triplet and sextet defined by $SU(3)$ symmetry are not the  physical states. A physical singly heavy baryon state must be a mixture of the corresponding anti-triplet and sextet states.

In the study of semi-leptonic charmed baryons decays, recent experimental measurements on the decay width by BESIII and Belle collaborations imply significant flavor $SU(3)$ symmetry breaking \cite{BESIII:2015ysy,Belle:2021crz,Belle:2021dgc,BESIII:2023vfi}, while a recent Lattice QCD calculation of transition form factors finds less severe symmetry breaking~\cite{Farrell:2023vnm}. 
To understand this phenomenon, various possible mechanisms were explored in Ref.\cite{He:2021qnc}, with a very compelling contender being the incorporation of $\Xi_c$-$\Xi_c^{\prime}$ mixing~\cite{Geng:2022yxb}. It was found that to realize the sizable flavor $SU(3)$ breaking in the semi-leptonic $\Xi_c$ decays, a large $\Xi_c-\Xi_c^{\prime}$ mixing angle is required: $\theta=24.66^{\circ}\pm 0.90^{\circ}$ \cite{Geng:2022yxb,Geng:2022xfz} and $\theta=16.27^{\circ}\pm 2.30^{\circ}$ \cite{Ke:2022gxm}. Based on these observations, a method to measure this mixing angle in a four-body $\Xi_c$ decay has been proposed by Ref.~\cite{Xing:2022phq}.   On the other hand, direct calculation using QCD sum rules  gives $\theta=5.5^{\circ}\pm 1.8^{\circ}$ \cite{Aliev:2010ra} and $\theta=2.0^{\circ}\pm 0.8^{\circ}$ \cite{Sun:2023noo};  A recent lattice QCD calculation gives $\theta=1.2^{\circ}\pm 0.1^{\circ}$ \cite{Liu:2023feb}, confirmed by  an improved determination~\cite{Liu:2023pwr}. Heavy quark effective theory gives $\theta=8.12^{\circ}\pm 0.80^{\circ}$ \cite{Matsui:2020wcc}. 

%Recently, an improved method to determine the $\Xi_c$-$\Xi_c^{\prime}$ mixing angle has been proposed, where besides QCD, the flavor $SU(3)$ breaking effect is 
%introduced to correct the mixing angle \cite{Liu:2023pwr}. 
%In this literature, the flavor $SU(3)$ breaking is mainly from the $s, u/d$ mass difference. A lattice QCD calculation is performed to extract the mixing angle from the matrix element of the $SU(3)$ breaking Hamiltonian.
%On the other hand, the electric charge difference between the $u,d,s$ quarks can also lead to the $SU(3)$ breaking, where the QED effect cannot be neglected.

In this work, we will concentrate on the QED corrections for the $\Xi_c$-$\Xi_c^{\prime}$ mixing which have not been taken into account in the previous investigations. Unlike QCD, the QED contributions cannot be directly calculated  on the lattice. Therefore, we will employ the light-front quark model (LFQM) in the three-quark picture~\cite{Zhao:2023yuk, Tawfiq:1998nk, Cheng:2003sm, Cheng:1996if} to evaluate the corresponding matrix element at the leading order in $\alpha_{\rm EM}$.  We will also explore the dependence of the mixing angle on the LFQM parameters to ensure the result is stable.

The rest of this paper is organized as follows. In Sec.~\ref{sec:mixing_angle},
we will introduce the $SU(3)$ breaking Lagrangian, and construct the necessary matrix elements of the $SU(3)$ breaking Hamiltonian for extracting the $\Xi_c$-$\Xi_c^{\prime}$ mixing angle. 
An introduction to  the three-quark light-front quark model and the calculation of QED contributions will be given in Sec.~\ref{sec:LFQM} and Sec.~\ref{sec:formulae_angle}.
Numerical results for $\Xi_c$-$\Xi_c^{\prime}$ mixing angle and the dependence on the input parameters are given in Sec.~\ref{sec:num_result}. The last paragraph contains  a brief summary of this work.

\section{The $\Xi_c$-$\Xi_c^{\prime}$ mixing}
\label{sec:mixing_angle}

The flavor SU(3) symmetry is very convenient to analyse the weak decay amplitudes and for some applications, one can see Refs.~\cite{Zeppenfeld:1980ex,Savage:1989ub,Deshpande:1994ii,Chau:1986du,Chau:1990ay,Lu:2016ogy,Wang:2017azm,He:2018php}. Despite of its success, 
the flavor $SU(3)$ symmetry is an approximate symmetry which are broken by the $u, d, s$ quark  mass differences and QED contributions. In the following we  assume the same mass for up and down quarks, and accordingly the full QCD+QED Lagrangian contains both the terms conserving and breaking the flavor $SU(3)$ symmetry: $\mathcal{L}_{\rm QCD+QED}=\mathcal{L}_0+\Delta \mathcal{L}$. The $SU(3)$ conserving term $\mathcal{L}_0$ reads as
\begin{eqnarray}
\mathcal{L}_0&=&\sum_{q}\bar{\psi}_q(i\slashed{D}-m_u)\psi_q+e\sum_{q}e_s\bar{\psi}_q\slashed{A}\psi_q\nonumber\\
&+&ee_c\bar{\psi}_c\slashed{A}\psi_c,
\end{eqnarray}
where $D$ is the covariant derivative of QCD. $q=u,d,s$, $e_q$ is the electric charge of $q$ and $m_u=m_d=m_s$ is assumed in the flavor $SU(3)$ symmetry. The $SU(3)$ symmetry breaking term $\Delta \mathcal{L}$ arises from the mass difference  and charge difference  the $s$ quark and the $d$ and/or $u$ quarks. Explicitly, $\Delta \mathcal{L}$ can be divided into a mass term and a charge term:
\begin{eqnarray}
\Delta \mathcal{L}&=&\bar{\psi}_s(m_u-m_s)\psi_s+e(e_u-e_s)\bar{\psi}_u\slashed{A}\psi_u.
\end{eqnarray}
Similarly, the Hamiltonian is decomposed as:
\begin{eqnarray}
  H=H_0+\Delta H,   
\end{eqnarray}
with 
\begin{eqnarray}
 \Delta H = \int d^3x \Delta \mathcal{H}(x) = - \int d^3x \Delta \mathcal{L}(x). 
\end{eqnarray}

The $\Xi_c$ baryons are composed of $c,u,s$ quarks. In the $SU(3)$ sysmmetry limit, they are classified into an anti-triplet $\Xi_c^{\bar 3}$ and a sextet $\Xi_c^{6}$, which are also eigenstates of the $SU(3)$ conserved Hamiltonian $H_0$. The corresponding eigenstates are defined as
\begin{eqnarray}
H_0 |\Xi_c^{\bar{3}}\rangle=m_{\Xi_c^{\bar{3}}}|\Xi_c^{\bar{3}}\rangle,\ \ 
H_0|\Xi_c^{6}\rangle=m_{\Xi_c^{6}}|\Xi_c^{6}\rangle.
\end{eqnarray}
On the other hand, the full Hamiltonian is diaogonalized by the two physical mass eigenstates: 
\begin{eqnarray}
H |\Xi_c\rangle=m_{\Xi_c}|\Xi_c\rangle,\ \ 
H |\Xi_c^{\prime}\rangle=m_{\Xi_c^{\prime}}|\Xi_c^{\prime}\rangle.\label{eq:physicalMassDef}
\end{eqnarray}
The mixing between the physical doublet $|P\rangle=(|\Xi_c\rangle, |\Xi_c^{\prime}\rangle)^T$ and the $SU(3)$ doublet $|S\rangle=(|\Xi_c^{\bar{3}}\rangle, |\Xi_c^{6}\rangle)^T$ is expressed by a unitary transforming matrix $U$ parameterized by a mixing angle $\theta$:
\begin{eqnarray}
|P\rangle=
\begin{pmatrix}
\cos\theta & \sin\theta\\
-\sin\theta & \cos\theta
\end{pmatrix}
|S\rangle
=U|S\rangle.\label{eq:unitaryTransfer}
\end{eqnarray}

Now we consider the matrix element for the $SU(3)$ doublet $|S\rangle=(|\Xi_c^{\bar{3}}(P_{\bar 3})\rangle, |\Xi_c^{6}(P_{6})\rangle)^T$: 
$\langle S|H|S\rangle$, and set both the intial and final states to be static ${\vec P}={\vec P}^{\prime}=0$ and on-shell $P_{\bar 3}^0=m_{\Xi_c^{\bar 3}}, P_{6}^0=m_{\Xi_c^6}$.  Using the the unitary transformation $U$ defined in Eq.~\eqref{eq:unitaryTransfer} as well as the physical masses defined in Eq.~\eqref{eq:physicalMassDef}, we have
\begin{align}
&\begin{pmatrix}
\langle \Xi_c^{\bar{3}}(S_z')|H|\Xi_c^{\bar{3}}(S_z)\rangle &\langle \Xi_c^{6}(S_z')|H|\Xi_c^{\bar{3}}(S_z)\rangle\\
\langle \Xi_c^{\bar{3}}(S_z')|H|\Xi_c^{6}(S_z)\rangle & \langle \Xi_c^{6}(S_z')|H|\Xi_c^{6}(S_z)\rangle
\end{pmatrix}\nonumber\\
=& 2 (2\pi)^3\delta^{(3)}(\vec{0})\delta_{S_zS_z'}\nonumber\\
&\times \begin{pmatrix}
m_{\Xi_c}^2 \cos^2\theta+m_{\Xi_c'}^2\sin^2\theta &(m_{\Xi_c}^2 -m_{\Xi_c'}^2)\cos\theta \sin\theta\\
(m_{\Xi_c}^2 -m_{\Xi_c'}^2)\cos\theta \sin\theta & m_{\Xi_c}^2 \sin^2\theta+m_{\Xi_c'}^2 \cos^2\theta 
\end{pmatrix},\label{eq:SHSmatrix}\nonumber\\
\end{align}
where the momentum dependence  of the $\Xi_c^{\bar 3, 6}$ states are not shown.

\begin{comment}
\begin{eqnarray}
&&\begin{pmatrix}
\langle \Xi_c^{\bar{3}}(\vec{P}',S_z')|H|\Xi_c^{\bar{3}}(\vec{P},S_z)\rangle &\langle \Xi_c^{6}(\vec{P}',S_z')|H|\Xi_c^{\bar{3}}(\vec{P},S_z)\rangle\\
\langle \Xi_c^{\bar{3}}(\vec{P}',S_z')|H|\Xi_c^{6}(\vec{P},S_z)\rangle & \langle \Xi_c^{6}(\vec{P}',S_z')|H|\Xi_c^{6}(\vec{P},S_z)\rangle
\end{pmatrix}|_{\vec{P}=0}  \nonumber\\
&= & 2 (2\pi)^3\delta^{(3)}(\vec{P}'-\vec{P})\delta_{S_zS_z'}\nonumber\\
&&\times \begin{pmatrix}
m_{\Xi_c}^2 \cos^2\theta+m_{\Xi_c'}^2\sin^2\theta &(m_{\Xi_c}^2 -m_{\Xi_c'}^2)\cos\theta \sin\theta\\
(m_{\Xi_c}^2 -m_{\Xi_c'}^2)\cos\theta \sin\theta & m_{\Xi_c}^2 \sin^2\theta+m_{\Xi_c'}^2 \cos^2\theta 
\end{pmatrix}|_{\vec{P}=0}.\label{eq:SHSmatrix}\nonumber\\
\end{eqnarray}
\end{comment}

Choosing the upper-right off-diagonal component in the Eq.~\eqref{eq:SHSmatrix}, we have
\begin{eqnarray}
&&\langle \Xi_c^{6}(S_z')|H|\Xi_c^{\bar{3}}(S_z)\rangle\nonumber\\
&=&(2\pi)^3\delta^{(3)}(\vec{0})\delta_{S_zS_z'}(m_{\Xi_c}^2 -m_{\Xi_c'}^2)\sin 2\theta.\label{eq:3bar6matrix}
\end{eqnarray}
To extract the mixing angle $\theta$, one has to calculate the matrix element on the left hand side above, which can be further expressed as
\begin{eqnarray}
&&\langle \Xi_c^{6}(S_z')|H|\Xi_c^{\bar{3}}(S_z)\rangle\nonumber\\
&=& (2\pi)^3\delta^{(3)}({\vec{0}})\langle \Xi_c^{6}(S_z')|\Delta \mathcal{H}(0)|\Xi_c^{\bar{3}}(S_z)\rangle.\label{eq:3bar6matrix2}
\end{eqnarray}
Relating Eq.~\eqref{eq:3bar6matrix} and Eq.~\eqref{eq:3bar6matrix2} we have
\begin{eqnarray}
\frac{\langle \Xi_c^{6}(S_z')|\Delta \mathcal{H}(0)|\Xi_c^{\bar{3}}(S_z)\rangle}{m_{\Xi_c}^2 -m_{\Xi_c'}^2} = \delta_{S_zS_z'}\sin2\theta.
\end{eqnarray}
Since the mixing angle is independent of the baryon spin, we can average it and obtain
\begin{eqnarray}
\label{eq:mixing_angle}
&&\sin2\theta =\frac{1}{2}\sum_{S_z} \frac{\langle \Xi_c^{6}(S_z)|\Delta \mathcal{H}(0)|\Xi_c^{\bar{3}}(S_z)\rangle|}{m_{\Xi_c}^2 -m_{\Xi_c'}^2},\label{eq:MixangleFormula}
\end{eqnarray}
where the $1/2$ comes from the average of baryon spin.

\section{light-front quark model for baryon}
\label{sec:LFQM}

The mixing angle of $\Xi_c-\Xi_c^{\prime}$ can be extracted from  Eq.~\eqref{eq:MixangleFormula}. 
Here we will use LFQM in the three quark picture~\cite{Tawfiq:1998nk,Zhao:2023yuk} to calculate the matrix element on the right hand side of Eq.~\eqref{eq:MixangleFormula}. We will firstly introduce the three quark picture and list the necessary formulas for the calculation in this work. In LFQM a $\Xi_c^{\bar{3}/6}$ baryon state can be expressed as
\begin{widetext}
\begin{eqnarray}
\left|\Xi_c^{\bar{3}/6}\right\rangle 
&=& \int \Big(\Pi_{i=1}^3\left\{d^3 \tilde{p}_i\right\}\Big) 2(2 \pi)^3 \frac{\delta^{(3)}\left(\tilde{P}-\tilde{p}_1-\tilde{p}_2-\tilde{p}_3\right)}{\sqrt{P^{+}}}\sum_{\lambda_i} \Psi^{S,S_z}_{\bar{3}/6}\frac{\epsilon^{i j k}}{\sqrt{6}}\left|c^i\left(p_1, \lambda_1\right) s^j\left(p_2, \lambda_2\right) u^k\left(p_3, \lambda_3\right)\right\rangle,\label{eq:LFQMbaryonstate}
\end{eqnarray}
\end{widetext}
where $P$ denotes the baryon momentum, $p_i$ and $\lambda_i$ are the momentums and helicities of the constituent quarks.  $\Psi^{S,S_z}_{\bar{3}/6}$ denotes the spin and momentum wave function.

In the light-cone coordinates, the momentum and the corresponding integration measure are written as
\begin{eqnarray}
&&p =\left(p^{+}, p^{-}, p_{\perp}\right),  ~~\tilde{p}=\left(p^{+}, p_{\perp}\right),~~p^{ \pm}  =p^0 \pm p^3,\nonumber\\
&&\left\{d^3 \tilde{p}\right\} \equiv \frac{d p^{+} d^2 p_{\perp}}{2(2 \pi)^3\sqrt{p^+} },~~\frac{d^4 p}{(2 \pi)^3}  \equiv \frac{d p^{-} d p^{+} d^2 p_{\perp}}{2(2 \pi)^3}.
\end{eqnarray}
The three-momentum in the light-cone frame is defined as $\tilde{p}=(p^+,{\vec p}_{\perp})$. We introduce two intrinsic variables for the constituent quarks, the light-cone momentum fraction $x_i$ and the transversal momentum  $k_{i\perp}$:
\begin{eqnarray}
p_i^+=x_iP^+,~~{\vec p}_{i\perp}=x_i{\vec P}_\perp+{\vec k}_{i\perp},~~\sum_{i=1}^3 {\vec k}_{i\perp}=0,
\end{eqnarray}
where $0<x_i<1$, the constraints:  $\sum_{i=1}^3 x_i=1$ and $\sum_{i=1}^3 {\vec k}_{i\perp}=0$ are imposed due to the momentum conservation. The total momentum of the constituent quarks is denoted as $\bar P=\sum_{i=1}^3 p_i$. The invariant mass $M_0$ is defined as $M_0^2\equiv\bar{P}^2$. It should be noted that $\bar P$ is not equal to the baryon momentum $P$ since the momentum of  baryon and its constituent quarks can not be on-shell simultaneously.  Choosing a frame where  $P_\perp=0$, we can express the invariant mass $M_0$ as
\begin{eqnarray}
M_0^2=\sum_{i=1}^3\frac{{\vec k}_{i\perp}^2+m_i^2}{x_i}.
\end{eqnarray}

The internal momentum of the constituent quarks is defined as
\begin{eqnarray}
k_i=(k_i^+,k_i^-,k_{i\perp})=(x_iM_0,\frac{{\vec k}_{i\perp}^2+m_i^2}{x_iM_0},k_{i\perp}).
\end{eqnarray}
Then in the Cartesian coordinate the components of $k_i=(e_i,{\vec k}_{i\perp},k_{iz})$ can be written as
\begin{eqnarray}
e_i=\frac{k_i^++k_i^-}{2}=\frac{x_iM_0}{2}+\frac{{\vec k}_{i\perp}^2+m_i^2}{2x_iM_0},\nonumber\\
k_{iz}=\frac{k_i^+-k_i^-}{2}=\frac{x_iM_0}{2}-\frac{{\vec k}_{i\perp}^2+m_i^2}{2x_iM_0}.
\end{eqnarray}

The wave function $\Psi^{S,S_z}$ for the anti-triplet state $\Xi_c^{\bar{3}}$ has the form of
\begin{eqnarray}
\Psi_{\bar{3}}&=&A\bar{u}_{\lambda_3}\left(p_3\right)\left(\slashed{\bar{P}}+M_0\right)\left(-\gamma_5\right) C \bar{u}_{\lambda_2}^T\left(p_2\right) \nonumber\\
&&\times\bar{u}_{\lambda_1}\left(p_1\right) u\left(\bar{P}\right)\Phi(x_i,k_{i\perp}),\label{eq:3barWF}
\end{eqnarray}
where the $u, s$ quark form a $0^+$ diquark. The wave function  for the anti-triplet state $\Xi_c^{6}$ has the form 
\begin{eqnarray}
\Psi_{6}&=&A\bar{u}_{\lambda_3}\left(p_3\right)\left(\slashed{\bar{P}}^{\prime}+M_0^{\prime}\right)\left(\gamma^\mu-v^\mu\right) C \bar{u}_{\lambda_2}^T\left(p_2\right) \nonumber\\
&&\times\bar{u}_{\lambda_1}\left(p_1\right)\left(\frac{1}{\sqrt{3}} \gamma_\mu \gamma_5\right) u\left(\bar{P}^{\prime}\right)\Phi(x_i,k_{i\perp}^{\prime}),\label{eq:6WF}
\end{eqnarray}
where the $u, s$ quarka form a $1^+$ diquark, and $v^\mu=\bar{P}^{\prime \mu}/M_0^{\prime}$. The three-particle  momentum wave function $\Phi$ describes the relative motion between two of the three constituent quarks as well as the relative motion between the third one and the center of other two quarks. Their explicit expressions are 
\begin{eqnarray}
\label{eq:wave_function_shape}
\Phi(x_i,k_{i\perp})&=&\sqrt{\frac{e_1e_2e_3}{x_1x_2x_3M_0}}\phi(\vec{k}_1,\beta_1)\phi(\frac{\vec{k}_2-\vec{k}_3}{2},\beta_{23}),\nonumber\\
\phi(\vec{k},\beta)&=&4\left(\frac{\pi}{\beta^2}\right)^{\frac{3}{4}}e^{\frac{-k_\perp^2-k_z^2}{2\beta^2}},
\end{eqnarray}
where $\beta_1$ and $\beta_{23}$ are the shape parameters. The normalization factor $A$ is given as
\begin{eqnarray}
A=\frac{1}{4\sqrt{M_0^3(e_1+m_1)(e_2+m_2)(e_3+m_3)}},
\end{eqnarray}
where $m_1$, $m_2$ and $m_3$ represent the masses of the charm quark, the strange quark, and the up quark in Eq.~\eqref{eq:LFQMbaryonstate}, respectively.
\section{The QED contributions to the $\Xi_c-\Xi_c'$ mixing}
\label{sec:formulae_angle}

\begin{figure*}
\centering
\includegraphics[width=1.0\textwidth]{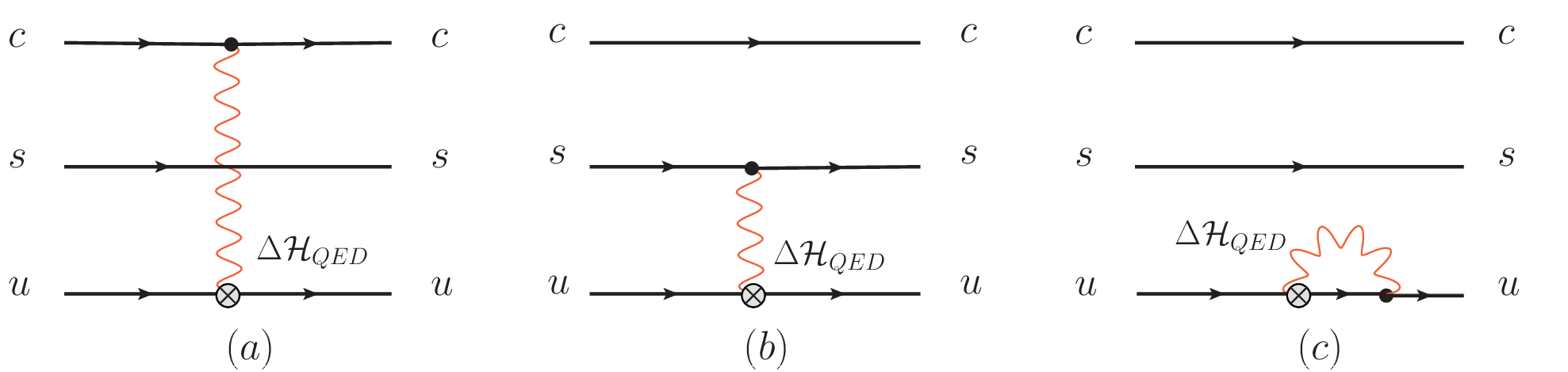}
\caption{Three leading-order diagrams for the matrix element in Eq.~\eqref{eq:DeltaHmatrix}. Only the diagram (a) breaks the heavy quark spin symmetry.}
\label{fig:heavy-quark-symmetry breaking}
\end{figure*}

\begin{figure}[h]
\centering
\includegraphics[width=0.8\textwidth]{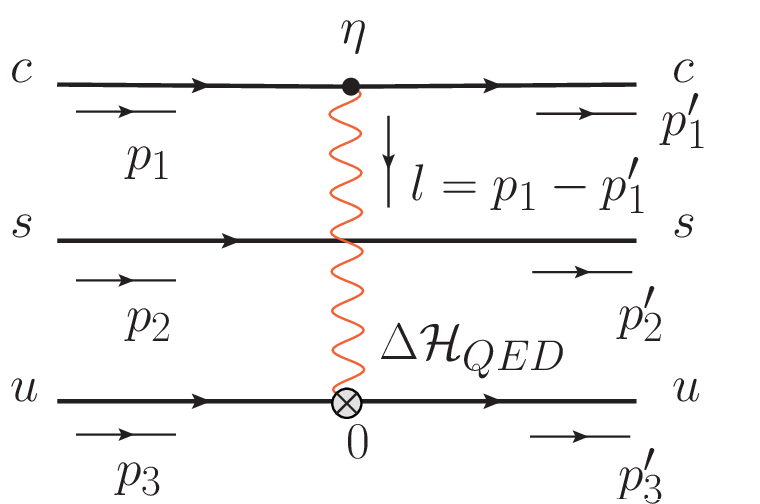}
\caption{The leading-order diagram that breaks the heavy quark spin symmetry for the matrix element in Eq.~\eqref{eq:DeltaHmatrix}.}
\label{fig:heavy-quark-symmetry breaking2}
\end{figure}

In this section, we calculate the QED effects on the $\Xi_c-\Xi_{c}^{\prime}$ mixing angle.
According to  Eq.~\eqref{eq:MixangleFormula}, we need to consider the matrix element
\begin{eqnarray}
\frac{1}{2}\sum_{S_z} \langle \Xi_c^{6}(P, S_z)|\Delta \mathcal{H}_{\rm QED}(0)|\Xi_c^{\bar{3}}(P, S_z)\rangle\label{eq:DeltaHmatrix}
\end{eqnarray}
with $\Delta \mathcal{H}_{\rm QED}=-e(e_u-e_s)\bar{\psi}_u\slashed{A}\psi_u$. 
The calculation is performed by using the light front baryon state given in Eq.~\eqref{eq:LFQMbaryonstate}, as well as the explicit expressions of the wave functions given in Eq.~\eqref{eq:3barWF} and \eqref{eq:6WF}. The corresponding Feynman diagrams are shown in Fig. \ref{fig:heavy-quark-symmetry breaking}, where the photon emitted from the $\Delta \mathcal{H}_{\rm QED}$ vertex attaches on the constituent quark lines.

It should be noticed that the (b),(c) diagrams in Fig. \ref{fig:heavy-quark-symmetry breaking} where no photon attaches on the charm quark are zero. This can be seen from the amplitude contributed from the charm quark line:
\begin{eqnarray}
&&\sum_{S_z}\sum_{\lambda_1} \bar{u}_{\lambda_1}(p_1)u_{S_z}(\bar{P})\bar{u}_{S_z^{\prime}}(\bar{P}) \left(\frac{\gamma_\mu\gamma_5}{\sqrt{3}}\right) u_{\lambda_1}(p_1^{\prime})\nonumber\\
&=&{\rm Tr}\left[ (\slashed{p}_1+m_1)(\bar{\slashed{P}}+M_0)  \left(\frac{1}{\sqrt{3}}\gamma_\mu\gamma_5\right) \right]=0,
\end{eqnarray}
which vanishes when the initial and final baryon spin are averaged. Note that the photon emission from the charm quark as shown in diagram (a) can change the charm quark spin, which breaks the heavy quark spin symmetry. In heavy quark effective theory, at the leading power the heavy quark spin is conserved, and the breaking effect occurs at the ${\cal O}(1/m_Q)$. Therefore, the QED induced $\Xi_c-\Xi_c^{\prime}$ mixing should be at the order of $1/m_c$. This also applies to the contributions from the scalar operators $\bar ss$.

Now the only contributing diagram  is shown in Fig.~\ref{fig:heavy-quark-symmetry breaking2}. In LFQM with three-quark picture its amplitude can be calculated as
\begin{eqnarray}
&&e(e_u-e_s) \left \langle \Xi_{c}(1)\right| \bar \psi_u(0) \slashed {A}(0)\psi_u(0)\left|\Xi_{c}(0)\right\rangle \nonumber\\
&=&-Q_c e^2 \int \frac{d x_1 d^2 k_{1\perp}}{2(2 \pi)^3\sqrt{x_1 } } d x_2 d^2 k_{2\perp} \frac{1}{\sqrt{1-x_1-x_2} } \nonumber\\
&& \times \int \frac{ d x_1^{\prime} d^2 k^{\prime }_{1\perp}}{2(2 \pi)^3\sqrt{x_1^{\prime}}} \frac{A^{\prime} A}{2(2 \pi)^3\sqrt{(1-x_2-x_1^{\prime})}} \nonumber\\
&& \times  \frac{\Phi(x_i,k_{i\perp}) \Phi(x_i^{\prime},k_{i\perp}^{\prime})}{(p_1-p_1^{\prime})^2+i\epsilon}{\rm {Tr}}_A{\rm {Tr}}_B,\label{eq:3barto6Amp}
\end{eqnarray}
where $Q_c=+\frac{2}{3}$ denotes the electric charge of the charm
quark, and the trace terms ${\rm {Tr}}_A$ and ${\rm {Tr}}_B$ are
\begin{eqnarray}
{\rm {Tr}}_A &=& {\rm {Tr}}\Big[(\slashed{p}_3+m_3)\left(\bar{\slashed{P}}+M_0\right)\left(-\gamma_5\right)(\slashed{p}_2-m_2) \nonumber\\
&&\times(\gamma^{\mu}-v^{\mu})(\bar{\slashed{P}}^{\prime}+M^{\prime}_0) (\slashed{p}^{\prime}_3+m_3)\gamma^{\nu}\Big],\nonumber\\
{\rm {Tr}}_B&=&{\rm {Tr}}\Big[
(\bar{\slashed{P}}+M_0)\frac{1}{\sqrt{3}} \gamma_\mu \gamma_5(\slashed{p}_1^{\prime}+m_1)\gamma_{\nu} (\slashed{p}_1+m_1)\Big].\nonumber\\
\end{eqnarray}
There are eight independent integration variables in Eq.~\eqref{eq:3barto6Amp}, which contains three transverse momentum module: $k_1=|{\vec k}_{1\perp}|, k_2=|{\vec k}_{2\perp}|, k_1^{\prime}=|{\vec k}_{1\perp}^{\prime}|$; three plus component momentum fractions: $x_1=k_1^{+}/P^{+}, x_2=k_2^{+}/P^{+}, x_1^{\prime}=k_1^{\prime+}/P^{+}$; The angle between ${\vec k}_{1\perp}$ and ${\vec k}_{2\perp}$ is $\alpha$; The angle between the momentum ${\vec k}_{1\perp}$ and ${\vec k}_{1\perp}^{\prime}$ is $\beta$. Using those eight variables we can write the amplitude as
\begin{eqnarray}
&&e(e_u-e_s) \left \langle \Xi_{c}(1)\right| \bar \psi_u(0) \slashed {A}(0)\psi_u(0)\left|\Xi_{c}(0)\right\rangle|\nonumber\\
&=&-\frac{2 e^2}{3} \int_0^{1} d x_1 \int_0^{1-x_1}  d x_2  \int_0^{1-x_2} d x_1^{\prime} \int_0^{\infty}  \frac{d k_1  d k_2 d k_1^{\prime}}{\Big[2(2 \pi)^3\Big]^3} \nonumber\\
&&\times 2\pi \int_0^{2\pi} d\alpha d\beta \frac{ k_1 k_2 k_1^{\prime}A^2} {\sqrt{x_1x_1^{\prime}(1-x_1-x_2)(1-x_1^{\prime}-x_2)}} \nonumber\\
&&\times \Phi(x_i,k_{i\perp})\Phi(x_i^{\prime},k_{i\perp}^{\prime}) \frac{{\rm {Tr}}_A{\rm {Tr}}_B}{l^2+i\epsilon},\label{eq:3barto6MatrixResult}
\end{eqnarray}
where $l$ is the photon momentum and $l^2=2m_1^2-2p_1\cdot p_1'$.

It should be mentioned that the above integration is infrared (IR) and ultraviolet (UV) finite. In the LFQM calculation all the internal quark lines as shown in Fig. \ref{fig:heavy-quark-symmetry breaking2} are on shell. Therefore the photon emitted from an on shell quark cannot be on shell $l^2=0$ unless its momentum vanishes $l=0$. Interestingly it can be found that the trace term in the nominator: ${\rm {Tr}}_A{\rm {Tr}}_B$ is proportional to quadratic forms of $l$ such as $l^2$ or $l^{\mu}l^{\nu}$, which cancels the singularity from the denominator so that prevents the IR divergence in the integration. Since the exponent of the wave function Eq.~\eqref{eq:wave_function_shape} depresses the contribution which from large momentum, there is also no UV divergence.

\section{Numerical results}
\label{sec:num_result}

%%%%%%%%%%%%%%%%%%%%%
\begin{figure*}[htb!]
\centering
\includegraphics[width=0.4\textwidth]{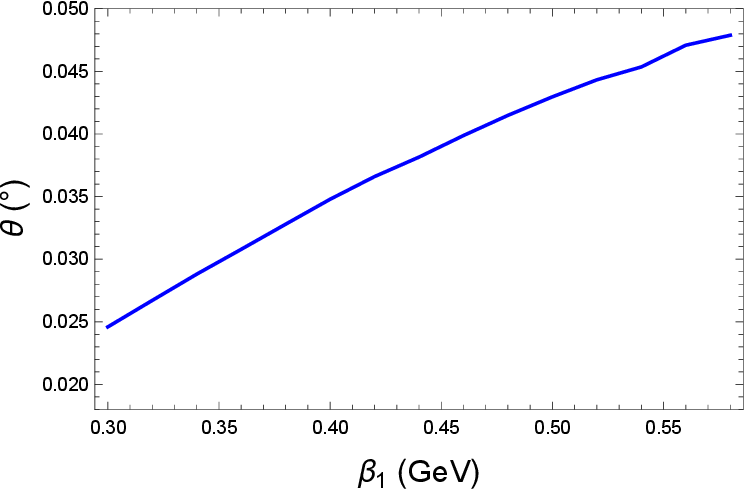}
\includegraphics[width=0.4\textwidth]{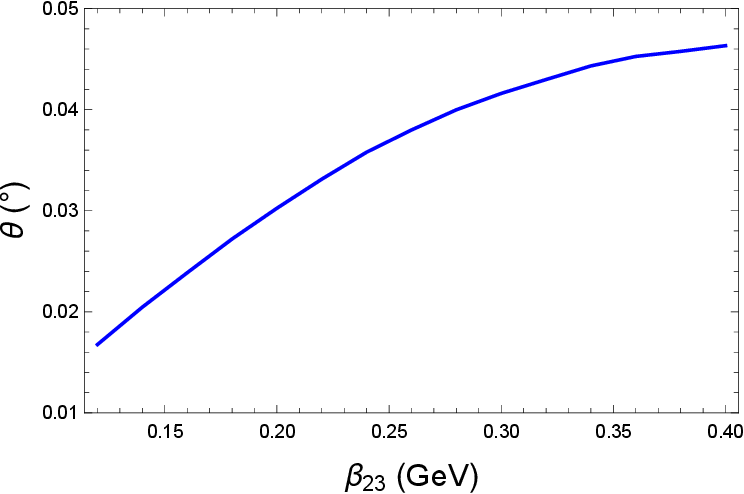}
\includegraphics[width=0.4\textwidth]{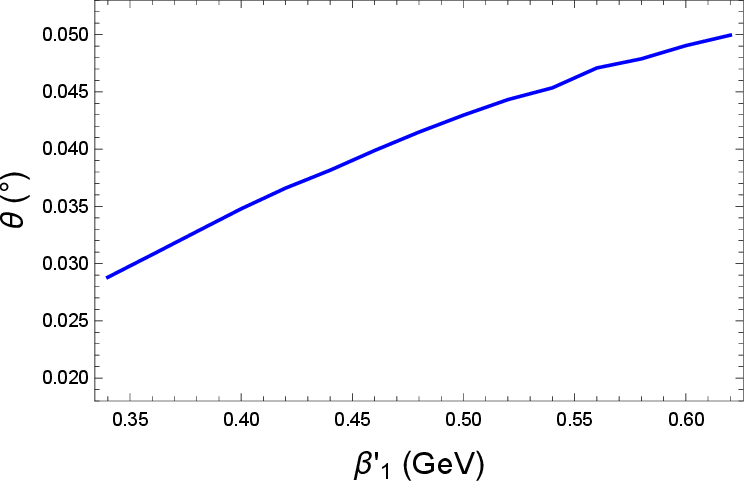}
\includegraphics[width=0.4\textwidth]{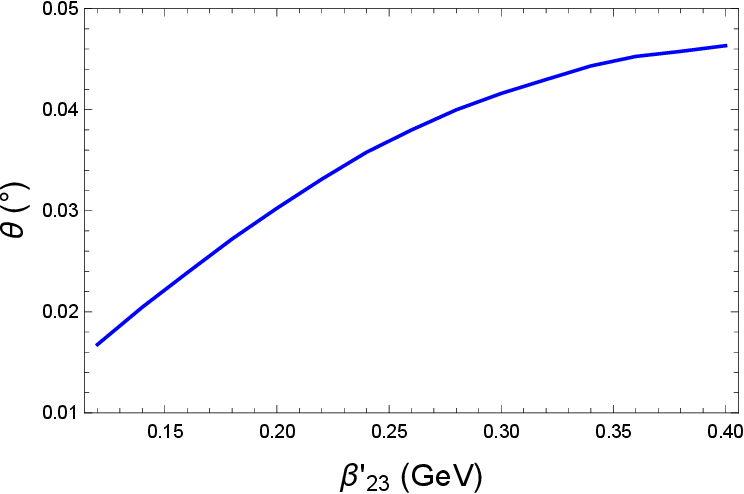}
\caption{The dependence of the mixing angle on the shape parameters.}
\label{fig:shap-parameter-dependence}
\end{figure*}
\begin{figure*}[htb!]
\centering
\includegraphics[width=0.40\textwidth]{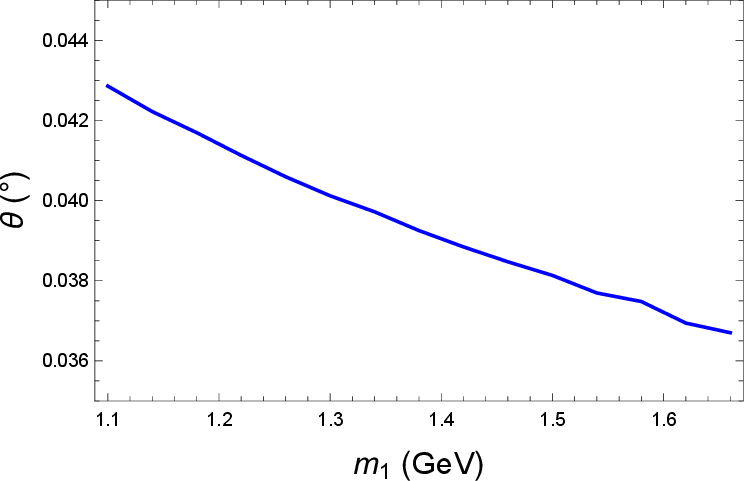}
\includegraphics[width=0.48\textwidth]{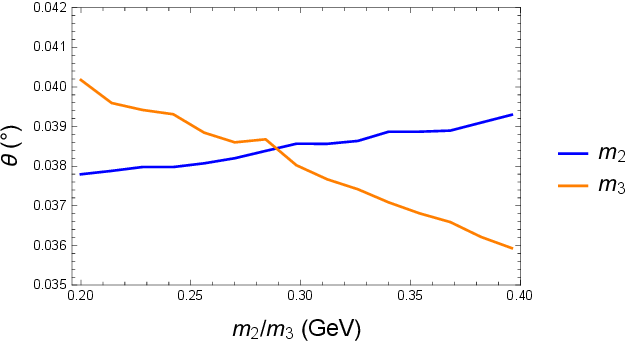}
\caption{The dependence of the mixing angle on quark mass.}
\label{fig:quark-mass-dependence}
\end{figure*}
%%%%%%%%%%%%%%%%%%%%%

In this work we will adopt the following constituent quark mass parameters:
\begin{eqnarray}
m_u=0.25~{\rm{GeV}},~~m_s=0.37~{\rm{GeV}},~~m_c=1.4~\rm{GeV},\nonumber
\end{eqnarray}
which can be found from \cite{Wang:2008xt, Wang:2009mi}. The shape parameters are extracted from  \cite{Zhao:2023yuk}:
\begin{eqnarray}
\beta_1^{\bar{3}}=0.45\pm0.05~\rm{GeV},~~\beta_{23}^{\bar{3}}=0.27\pm0.03~\rm{GeV},\nonumber\\
\beta_1^{6}=0.49\pm0.04~\rm{GeV},~~\beta_{23}^{6}=0.28\pm0.03~\rm{GeV}.
\end{eqnarray}
The baryon masses are taken as \cite{ParticleDataGroup:2022pth, Geng:2022yxb}
\begin{eqnarray}
m_{\Xi_c}=2.47~\rm{GeV},\  m_{\Xi_c^\prime}=2.58~\rm{GeV}.
\end{eqnarray}
According to Eq.~\eqref{eq:MixangleFormula} and Eq.~\eqref{eq:3barto6MatrixResult}, the mixing angle from  QED correction is obtained as
\begin{eqnarray}
\theta_{\rm QED} \approx 0.04^\circ.
\end{eqnarray}

%It can be found that the QED correction to the mixing angle is two orders of magnitude smaller than that of QCD correction~\cite{Liu:2023feb,Liu:2023pwr}. 

We  have also investigated the dependence of $\theta$ on the input parameters: the  shape parameters and the constituent heavy quark mass. The mixing angle as a functions of shape parameters are shown in Fig.~\ref{fig:shap-parameter-dependence}, while the mixing angle as a function of heavy quark mass are shown in Fig.~\ref{fig:quark-mass-dependence}. 

It should be noted that our result is  smaller than the contributions induced by the mass difference from the most recent lattice QCD calculation~\cite{Liu:2023pwr}.  The sum of these two kinds of contributions is still  smaller than the experimental measurement and cannot explain the large $SU(3)$ symmetry breaking in experiment.

\section{Conclusion}\label{conslusion}

%In this work, we consider the mass correction and QED correction effects of mixing angle. We have found that the mass correction of the mixing angle is suppressed. At the same time, we also find that in all Feynman diagrams where the heavy quark symmetry is not broken, the Feynman diagram does not contribute to the mixing angle, so the lowest mass correction is zero. The mixing of the mass correction term and the QED correction is depressed relative to the pure QED correction in the heavy quark approximation. Thus, in heavy quark effective field theory, the correction for the mixing angle is zero. 

In this work we have calculated the QED contribution to the $\Xi_c-\Xi_c'$ mixing which has not been taken into account in the previous analyses. 
In the calculation, we have employed the light-front quark model. We have explicitly demonstrated that the mixing breaks the heavy quark symmetry. 

Numerically, the QED contribution to the mixing angle is found about 0.04$^\circ$, and the result is less sensitive to the quark masses and shape parameters. Our result is smaller than the contributions induced by the mass difference from the most recent lattice QCD.  The sum of these two kinds of contributions is much smaller than the experimental measurement and cannot explain the large $SU(3)$ symmetry breaking in experiment.

\section*{Acknowledgement}

We thank Zhi-Peng Xing for valuable discussions, and Hang Liu, and Qi-An Zhang for useful discussions. This work is supported in part by Natural Science Foundation of China under grant No.U2032102, 12147140,12125503, 12061131006, 
12205180, 12305103, and 12335003. J.Z. is also partially supported by the Project funded by China Postdoctoral Science Foundation under Grant No. 2022M712088.


\begin{thebibliography}{}
%\cite{Zeppenfeld:1980ex}
\bibitem{Zeppenfeld:1980ex}
D.~Zeppenfeld,
%``SU(3) Relations for B Meson Decays,''
Z. Phys. C \textbf{8}, 77 (1981)
doi:10.1007/BF01429835
%214 citations counted in INSPIRE as of 01 Sep 2023

%\cite{Savage:1989ub}
\bibitem{Savage:1989ub}
M.~J.~Savage and M.~B.~Wise,
%``SU(3) Predictions for Nonleptonic B Meson Decays,''
Phys. Rev. D \textbf{39}, 3346 (1989)
[erratum: Phys. Rev. D \textbf{40}, 3127 (1989)]
doi:10.1103/PhysRevD.39.3346
%194 citations counted in INSPIRE as of 18 Sep 2023

%\cite{Deshpande:1994ii}
\bibitem{Deshpande:1994ii}
N.~G.~Deshpande and X.~G.~He,
%``CP asymmetry relations between anti-b0 ---\ensuremath{>} pi pi and anti-b0 ---\ensuremath{>} pi K rates,''
Phys. Rev. Lett. \textbf{75}, 1703-1706 (1995)
doi:10.1103/PhysRevLett.75.1703
[arXiv:hep-ph/9412393 [hep-ph]].
%96 citations counted in INSPIRE as of 15 Mar 2023

%\cite{He:1998rq}
\bibitem{He:1998rq}
X.~G.~He,
%``SU(3) analysis of annihilation contributions and CP violating relations in B ---\ensuremath{>} P P decays,''
Eur. Phys. J. C \textbf{9}, 443-448 (1999)
doi:10.1007/s100529900064
[arXiv:hep-ph/9810397 [hep-ph]].
%84 citations counted in INSPIRE as of 15 Mar 2023

%\cite{Chau:1987tk}
\bibitem{Chau:1987tk}
L.~L.~Chau and H.~Y.~Cheng,
%``Analysis of Exclusive Two-Body Decays of Charm Mesons Using the Quark Diagram Scheme,''
Phys. Rev. D \textbf{36}, 137 (1987)
doi:10.1103/PhysRevD.39.2788
%242 citations counted in INSPIRE as of 22 Sep 2023

%\cite{Gronau:1994rj}
\bibitem{Gronau:1994rj}
M.~Gronau, O.~F.~Hernandez, D.~London and J.~L.~Rosner,
%``Decays of B mesons to two light pseudoscalars,''
Phys. Rev. D \textbf{50}, 4529-4543 (1994)
doi:10.1103/PhysRevD.50.4529
[arXiv:hep-ph/9404283 [hep-ph]].
%443 citations counted in INSPIRE as of 01 Sep 2023

%\cite{Gronau:1995hm}
\bibitem{Gronau:1995hm}
M.~Gronau, O.~F.~Hernandez, D.~London and J.~L.~Rosner,
%``Broken SU(3) symmetry in two-body B decays,''
Phys. Rev. D \textbf{52}, 6356-6373 (1995)
doi:10.1103/PhysRevD.52.6356
[arXiv:hep-ph/9504326 [hep-ph]].
%247 citations counted in INSPIRE as of 26 Jul 2023

%\cite{Cheng:2014rfa}
\bibitem{Cheng:2014rfa}
H.~Y.~Cheng, C.~W.~Chiang and A.~L.~Kuo,
%``Updating B\textrightarrow{}PP,VP decays in the framework of flavor symmetry,''
Phys. Rev. D \textbf{91}, no.1, 014011 (2015)
doi:10.1103/PhysRevD.91.014011
[arXiv:1409.5026 [hep-ph]].
%82 citations counted in INSPIRE as of 03 Aug 2023

%\cite{Muller:2015lua}
\bibitem{Muller:2015lua}
S.~M\"uller, U.~Nierste and S.~Schacht,
%``Topological amplitudes in $D$ decays to two pseudoscalars: A global analysis with linear $SU(3)_F$ breaking,''
Phys. Rev. D \textbf{92}, no.1, 014004 (2015)
doi:10.1103/PhysRevD.92.014004
[arXiv:1503.06759 [hep-ph]].
%69 citations counted in INSPIRE as of 03 Aug 2023

%\cite{Shi:2017dto}
\bibitem{Shi:2017dto}
Y.~J.~Shi, W.~Wang, Y.~Xing and J.~Xu,
%``Weak Decays of Doubly Heavy Baryons: Multi-body Decay Channels,''
Eur. Phys. J. C \textbf{78}, no.1, 56 (2018)
doi:10.1140/epjc/s10052-018-5532-7
[arXiv:1712.03830 [hep-ph]].
%67 citations counted in INSPIRE as of 25 Sep 2023

%\cite{BESIII:2015ysy}
\bibitem{BESIII:2015ysy}
M.~Ablikim \textit{et al.} [BESIII],
%``Measurement of the absolute branching fraction for $\Lambda^+_{c}\to \Lambda e^+\nu_e$,''
Phys. Rev. Lett. \textbf{115}, no.22, 221805 (2015)
doi:10.1103/PhysRevLett.115.221805
[arXiv:1510.02610 [hep-ex]].
%97 citations counted in INSPIRE as of 25 Sep 2023

%\cite{Belle:2021crz}
\bibitem{Belle:2021crz}
Y.~B.~Li \textit{et al.} [Belle],
%``Measurements of the branching fractions of the semileptonic decays $\Xi_{c}^{0} \to \Xi^{-} \ell^{+} \nu_{\ell}$ and the asymmetry parameter of $\Xi_{c}^{0} \to \Xi^{-} \pi^{+}$,''
Phys. Rev. Lett. \textbf{127}, no.12, 121803 (2021)
doi:10.1103/PhysRevLett.127.121803
[arXiv:2103.06496 [hep-ex]].
%43 citations counted in INSPIRE as of 18 Sep 2023

%\cite{Belle:2021dgc}
\bibitem{Belle:2021dgc}
Y.~B.~Li \textit{et al.} [Belle],
%``First test of lepton flavor universality in the charmed baryon decays \ensuremath{\Omega}c0\textrightarrow{}\ensuremath{\Omega}-\ensuremath{\ell}+\ensuremath{\nu}\ensuremath{\ell} using data of the Belle experiment,''
Phys. Rev. D \textbf{105}, no.9, L091101 (2022)
doi:10.1103/PhysRevD.105.L091101
[arXiv:2112.10367 [hep-ex]].
%12 citations counted in INSPIRE as of 22 Sep 2023

%\cite{BESIII:2023vfi}
\bibitem{BESIII:2023vfi}
M.~Ablikim \textit{et al.} [BESIII],
%``Study of $\Lambda_c^+\rightarrow \Lambda \mu^+\nu_{\mu}$ and Test of Lepton Flavor Universality with $\Lambda_c^+\rightarrow \Lambda \ell^+\nu_{\ell}$ Decays,''
[arXiv:2306.02624 [hep-ex]].
%3 citations counted in INSPIRE as of 18 Sep 2023

%\cite{Farrell:2023vnm}
\bibitem{Farrell:2023vnm}
C.~Farrell and S.~Meinel,
%``Form factors for the charm-baryon semileptonic decay $\Xi_c\to \Xi \ell \nu$ from domain-wall lattice QCD,''
[arXiv:2309.08107 [hep-lat]].
%0 citations counted in INSPIRE as of 25 Sep 2023

%\cite{He:2021qnc}
\bibitem{He:2021qnc}
X.~G.~He, F.~Huang, W.~Wang and Z.~P.~Xing,
%``SU(3) symmetry and its breaking effects in semileptonic heavy baryon decays,''
Phys. Lett. B \textbf{823}, 136765 (2021)
doi:10.1016/j.physletb.2021.136765
[arXiv:2110.04179 [hep-ph]].
%20 citations counted in INSPIRE as of 18 Sep 2023

%\cite{Geng:2022yxb}
\bibitem{Geng:2022yxb}
C.~Q.~Geng, X.~N.~Jin and C.~W.~Liu,
%``Resolving puzzle in \ensuremath{\Xi}c0\textrightarrow{}\ensuremath{\Xi}\ensuremath{-}e+\ensuremath{\nu}e with \ensuremath{\Xi}c\ensuremath{-}\ensuremath{\Xi}c' mixing spectrum within the quark model,''
Phys. Lett. B \textbf{838}, 137736 (2023)
doi:10.1016/j.physletb.2023.137736
[arXiv:2210.07211 [hep-ph]].
%10 citations counted in INSPIRE as of 18 Sep 2023

%\cite{Geng:2022xfz}
\bibitem{Geng:2022xfz}
C.~Q.~Geng, X.~N.~Jin, C.~W.~Liu, X.~Yu and A.~W.~Zhou,
%``Semileptonic decays of doubly charmed baryons with \ensuremath{\Xi}c\ensuremath{-}\ensuremath{\Xi}c' mixing,''
Phys. Lett. B \textbf{839}, 137831 (2023)
doi:10.1016/j.physletb.2023.137831
[arXiv:2212.02971 [hep-ph]].
%6 citations counted in INSPIRE as of 18 Sep 2023

%\cite{Ke:2022gxm}
\bibitem{Ke:2022gxm}
H.~W.~Ke and X.~Q.~Li,
%``Revisiting the transition \ensuremath{\Xi}cc++\textrightarrow{}\ensuremath{\Xi}c(')+ to understand the data from LHCb,''
Phys. Rev. D \textbf{105}, no.9, 096011 (2022)
doi:10.1103/PhysRevD.105.096011
[arXiv:2203.10352 [hep-ph]].
%13 citations counted in INSPIRE as of 18 Sep 2023

%\cite{Xing:2022phq}
\bibitem{Xing:2022phq}
Z.~P.~Xing and Y.~j.~Shi,
%``Novel method for searching for the \ensuremath{\Xi}c0/+-\ensuremath{\Xi}c'0/+ mixing effect in the angular distribution analysis of a four-body \ensuremath{\Xi}c0/+ decay,''
Phys. Rev. D \textbf{107}, no.7, 074024 (2023)
doi:10.1103/PhysRevD.107.074024
[arXiv:2212.09003 [hep-ph]].
%4 citations counted in INSPIRE as of 22 Sep 2023

%\cite{Aliev:2010ra}
\bibitem{Aliev:2010ra}
T.~M.~Aliev, A.~Ozpineci and V.~Zamiralov,
%``Mixing Angle of Hadrons in QCD: A New View,''
Phys. Rev. D \textbf{83}, 016008 (2011)
doi:10.1103/PhysRevD.83.016008
[arXiv:1007.0814 [hep-ph]].
%26 citations counted in INSPIRE as of 12 Sep 2023

%\cite{Sun:2023noo}
\bibitem{Sun:2023noo}
X.~Y.~Sun, F.~W.~Zhang, Y.~J.~Shi and Z.~X.~Zhao,
%``Revisiting $\Xi_{Q}-\Xi_{Q}^{\prime}$ mixing in QCD sum rules,''
[arXiv:2305.08050 [hep-ph]].
%4 citations counted in INSPIRE as of 12 Sep 2023

%\cite{Liu:2023feb}
\bibitem{Liu:2023feb}
H.~Liu, L.~Liu, P.~Sun, W.~Sun, J.~X.~Tan, W.~Wang, Y.~B.~Yang and Q.~A.~Zhang,
%``\ensuremath{\Xi}c\ensuremath{-}\ensuremath{\Xi}c' mixing from lattice QCD,''
Phys. Lett. B \textbf{841}, 137941 (2023)
doi:10.1016/j.physletb.2023.137941
[arXiv:2303.17865 [hep-lat]].
%7 citations counted in INSPIRE as of 18 Sep 2023

%\cite{Liu:2023pwr}
\bibitem{Liu:2023pwr}
H.~Liu, W.~Wang and Q.~A.~Zhang,
%``An improved method to determine the $\Xi_c-\Xi_c'$ mixing,''
[arXiv:2309.05432 [hep-ph]].
%1 citations counted in INSPIRE as of 19 Sep 2023

%\cite{Matsui:2020wcc}
\bibitem{Matsui:2020wcc}
Y.~Matsui,
%``Mixing Angle of $\varXi_Q-\varXi_Q'$ in Heavy Quark Effective Therory,''
Nucl. Phys. A \textbf{1008}, 122139 (2021)
doi:10.1016/j.nuclphysa.2021.122139
[arXiv:2011.09653 [hep-ph]].
%4 citations counted in INSPIRE as of 16 May 2023

%\cite{Zhao:2023yuk}
\bibitem{Zhao:2023yuk}
Z.~X.~Zhao, F.~W.~Zhang, X.~H.~Hu and Y.~J.~Shi,
%``Baryons in the light-front approach: The three-quark picture,''
Phys. Rev. D \textbf{107}, no.11, 116025 (2023)
doi:10.1103/PhysRevD.107.116025
[arXiv:2304.07698 [hep-ph]].
%6 citations counted in INSPIRE as of 18 Sep 2023

%\cite{Tawfiq:1998nk}
\bibitem{Tawfiq:1998nk}
S.~Tawfiq, P.~J.~O'Donnell and J.~G.~Korner,
%``Charmed baryon strong coupling constants in a light front quark model,''
Phys. Rev. D \textbf{58}, 054010 (1998)
doi:10.1103/PhysRevD.58.054010
[arXiv:hep-ph/9803246 [hep-ph]].
%60 citations counted in INSPIRE as of 18 May 2023

%\cite{Cheng:2003sm}
\bibitem{Cheng:2003sm}
H.~Y.~Cheng, C.~K.~Chua and C.~W.~Hwang,
%``Covariant light front approach for s wave and p wave mesons: Its application to decay constants and form-factors,''
Phys. Rev. D \textbf{69}, 074025 (2004)
doi:10.1103/PhysRevD.69.074025
[arXiv:hep-ph/0310359 [hep-ph]].
%357 citations counted in INSPIRE as of 18 Sep 2023

%\cite{Cheng:1996if}
\bibitem{Cheng:1996if}
H.~Y.~Cheng, C.~Y.~Cheung and C.~W.~Hwang,
%``Mesonic form-factors and the Isgur-Wise function on the light front,''
Phys. Rev. D \textbf{55}, 1559-1577 (1997)
doi:10.1103/PhysRevD.55.1559
[arXiv:hep-ph/9607332 [hep-ph]].
%183 citations counted in INSPIRE as of 28 Jul 2023

%\cite{Chau:1986du}
\bibitem{Chau:1986du}
L.~L.~Chau and H.~Y.~Cheng,
%``Quark Diagram Analysis of Two-body Charm Decays,''
Phys. Rev. Lett. \textbf{56}, 1655-1658 (1986)
doi:10.1103/PhysRevLett.56.1655
%155 citations counted in INSPIRE as of 13 Sep 2023

%\cite{Chau:1990ay}
\bibitem{Chau:1990ay}
L.~L.~Chau, H.~Y.~Cheng, W.~K.~Sze, H.~Yao and B.~Tseng,
%``Charmless nonleptonic rare decays of $B$ mesons,''
Phys. Rev. D \textbf{43}, 2176-2192 (1991)
[erratum: Phys. Rev. D \textbf{58}, 019902 (1998)]
doi:10.1103/PhysRevD.43.2176
%293 citations counted in INSPIRE as of 18 Sep 2023

%\cite{Lu:2016ogy}
\bibitem{Lu:2016ogy}
C.~D.~L\"u, W.~Wang and F.~S.~Yu,
%``Test flavor SU(3) symmetry in exclusive $\Lambda_c$ decays,''
Phys. Rev. D \textbf{93}, no.5, 056008 (2016)
doi:10.1103/PhysRevD.93.056008
[arXiv:1601.04241 [hep-ph]].
%113 citations counted in INSPIRE as of 13 Sep 2023

%\cite{Wang:2017azm}
\bibitem{Wang:2017azm}
W.~Wang, Z.~P.~Xing and J.~Xu,
%``Weak Decays of Doubly Heavy Baryons: SU(3) Analysis,''
Eur. Phys. J. C \textbf{77}, no.11, 800 (2017)
doi:10.1140/epjc/s10052-017-5363-y
[arXiv:1707.06570 [hep-ph]].
%108 citations counted in INSPIRE as of 31 Aug 2023

%\cite{He:2018php}
\bibitem{He:2018php}
X.~G.~He and W.~Wang,
%``Flavor SU(3) Topological Diagram and Irreducible Representation Amplitudes for Heavy Meson Charmless Hadronic Decays: Mismatch and Equivalence,''
Chin. Phys. C \textbf{42}, no.10, 103108 (2018)
doi:10.1088/1674-1137/42/10/103108
[arXiv:1803.04227 [hep-ph]].
%35 citations counted in INSPIRE as of 15 Aug 2023

%\cite{Wang:2008xt}
\bibitem{Wang:2008xt}
W.~Wang, Y.~L.~Shen and C.~D.~Lu,
%``Covariant Light-Front Approach for B(c) transition form factors,''
Phys. Rev. D \textbf{79}, 054012 (2009)
doi:10.1103/PhysRevD.79.054012
[arXiv:0811.3748 [hep-ph]].
%144 citations counted in INSPIRE as of 18 Sep 2023

%\cite{Wang:2009mi}
\bibitem{Wang:2009mi}
X.~X.~Wang, W.~Wang and C.~D.~Lu,
%``B(c) to P-Wave Charmonia Transitions in Covariant Light-Front Approach,''
Phys. Rev. D \textbf{79}, 114018 (2009)
doi:10.1103/PhysRevD.79.114018
[arXiv:0901.1934 [hep-ph]].
%57 citations counted in INSPIRE as of 18 Sep 2023

%\cite{ParticleDataGroup:2022pth}
\bibitem{ParticleDataGroup:2022pth}
R.~L.~Workman \textit{et al.} [Particle Data Group],
%``Review of Particle Physics,''
PTEP \textbf{2022}, 083C01 (2022)
doi:10.1093/ptep/ptac097
%1662 citations counted in INSPIRE as of 25 Sep 2023
\end{thebibliography}
\end{document}